# What Have Spacetime, Shape and Symmetry to Do with Thermodynamics?

Jim McGovern

School of Mechanical and Transport Engineering,
Dublin Institute of Technology, Ireland, jim.mcgovern@dit.ie

ABSTRACT: Some highly speculative and serendipitous ideas that might relate thermodynamics, spacetime, shape and symmetry are brought together. A hypothetical spacetime comprising a pointwise 4-colour rhombohedral lattice with a fixed metric is considered. If there were absolute symmetry between distance and time it is suggested that the velocity of light would be an integer conversion factor between the units of these quantities. In the context of such outlandish hypotheses the answer to the question posed would be 'everything'.

## 1. INTRODUCTION

I have always found there to be some very disquieting terms and concepts in thermodynamics; these are topics at the end of short 'what-is?' or 'why?' chains starting with just about anything and ending with names such as force, mass, energy, temperature, volume, time, action, distance, direction. Generally as an engineer I got on with the tasks at hand. As a teacher I have told my students there are certain things that we don't understand, but take as given starting points from which we can develop and apply practical scientific and engineering techniques that allow us to understand to a significant degree how things work and therefrom to predict how different physical arrangements will behave and so solve engineering problems or come up with more-or-less ingenious or new designs and devices.

But the frustrating 'why questions' a little below the surface of what is scientifically understood have drawn me relentlessly to chase things that really should have been outside my domain as a Mechanical Engineer who, though in an academic post, is expected to be practically-minded.

For this paper on the above topic I have organised some of the notes that I have made, some of the references I have consulted and some of the things that I have tried out.

## 2. SPACETIME, SHAPE AND SYMMETRY

2.1 Spacetime

We live in a 3-D spatial world and in our simple experience and our natural reference frame we can be at the same place at different times so, in all, it seems that spacetime is 4-D. Therefore it seems natural that four co-ordinates should be required to describe the location of an event in spacetime.

Time and distance are fundamentally equivalent and indistinguishable. Time is embedded in distance and distance is embedded in time. The choice of the 4-D reference frame should be arbitrary. Physical reality is independent of any change in the orientation (i.e. rotation), just as it is independent of any change of the position (i.e. translation) of the reference frame. A general change of the position and orientation of the reference frame is called a spacetime transformation and involves changes in the coordinates of events with respect to the reference frame. It may also involve changes in any parameters that depend on those coordinates. Of course, certain parameters that depend on the spacetime coordinates may yet be invariant under spacetime transformations.

In spite of Einstein's realisations that mass and other parameters depend on velocity and that spacetime itself can be curved and distorted, there is a remarkable metrical consistency that can be observed throughout visible spacetime. Fundamental entities such as atoms, neutrinos, quarks or photons exist in innumerable quantities and 'measure' the same for each type. It also seems that Planck's constant, the Planck length and the Planck time are expressions of the universality of the metric.

Quantum field theory [1], quantum chromodynamics theory [2] and observations in particle physics must somehow be precisely consistent with cosmological observations and with Einstein's theories of special and general relativity. History shows that good theories evolve and become reconciled, thus revealing deeper beauty. An eventual 'theory of everything' will be expected to explain why the current theories do not entirely fit together over the huge range of orders of magnitude of the observations that are being made.

The absolute precision of nature's metrics points to a discrete nature of spacetime, 'the vacuum'. Atoms of the same type have the same size. Different single-atom clocks keep the same time. This suggests that spacetime is a lattice, based on a fundamental metric. There is a smallest quantum. This may well be the Planck distance (or time).

2.2 Shape

If spacetime is a point lattice with a metric then it has shape. Notwithstanding the distortions of spacetime when velocities and gravitation are involved, four arbitrary points in spacetime can form a line, a surface or a volume. Irrespective of the arbitrary number of points in discrete spacetime that might be selected arbitrarily, the selection has a describable shape. Of course any everyday object like a hat that exists in a region of spacetime has an immensely complex shape. This 'shape' includes all events of the hat for which it can, in principle, be identified as a hat.

2.3 Symmetry

Symmetry is sameness when objects are viewed at different spacetime coordinates or when an object is viewed from different spacetime coordinates. Hermann Weyl's book [3] on the topic included the statement of a general principle (p. 125) 'If conditions which uniquely determine their effect possess certain symmetries, then the effect will exhibit the same symmetry' and the remark (p. 126) 'As far as I can see, all a priori statements in physics have their origin in symmetry.'

Symmetry underlies quantum mechanics generally, quantum electrodynamics, quantum chromodynamics, the Standard Model of particle physics, general and special relativity and the observed structure of the universe at all scales.

2.4 Lineland, 1-D Spacetime; or is it 2-D?

Edwin Abbott [4] through his monograph 'Flatland' has stimulated much thought on the nature of dimensions and space. Thinking about a 'world' of only one dimension ('Lineland') or of two (Flatland) proves useful in appreciating the universe in which we actually find ourselves (which is normally thought of as having three 'space' dimensions). A key observation that can be made on consideration of Abbott's depiction of a hypothetical one-dimensional world is that an object or a pattern could not be inverted by means of a displacement alone. An object such as a lower case 'i' that points upwards in one-dimensional space could never become an object that points downwards by means of a displacement. Fig. 1(a) shows the letter 'i' and an identical letter displaced below the first. A reference point is shown mid-way between the two letters, but there is no symmetry about it. Figure 1(b) shows the letter 'i' and an upside-down letter 'i'. Again there is a reference point between them and there is symmetry about that point. Here the reference point is an 'axis of symmetry'.

Several issues arise from consideration of such one-dimensional cases. One of these is the number of dimensions: perhaps the up-direction and the down-direction constitute two dimensions, not one. If the up-direction and the down-direction are taken as two different (but collinear) dimensions, which each have a unique direction, then an upright 'i' in each dimension, as shown in Fig. 1(b), constitutes a symmetrical pair of the letter 'i'.

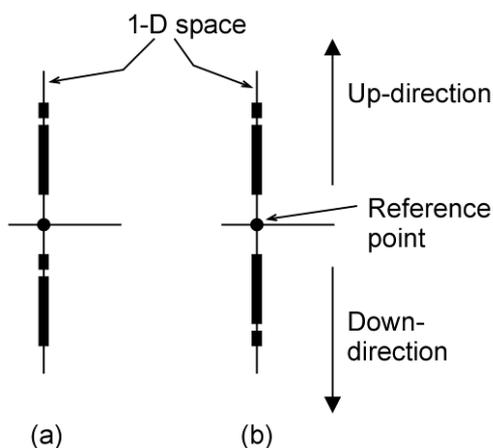

**Figure 1**: The letter 'i' in 1-D space (Lineland)

## 3. THE BELL FULLER LATTICE

Alexander Graham Bell devised a cuboctahedral truss made up of struts where each interior node was a joint or gusset for twelve struts [5], [6]. Exactly the same configuration of nodes was described by Richard Buckminster Fuller as the Isotropic Vector Matrix [7]. This configuration provides the basis of a volume-occupying lattice (conventionally described as a 3-D lattice). Fig. 2 illustrates an interior node and twelve nodes that surround it.

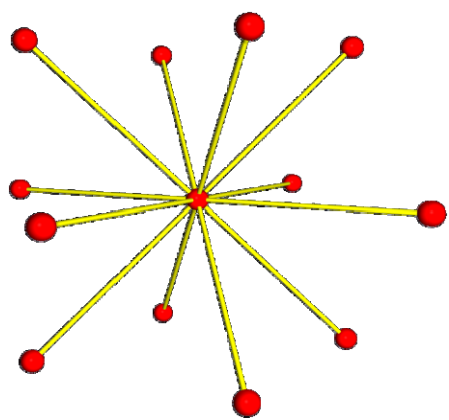

**Figure 2**: Node configuration of the Bell Fuller lattice.

The Bell Fuller lattice formed by replication of the node pattern shown in Fig. 2 has quite amazing symmetry properties. For instance, it includes all three of the only planar configurations of discrete points that are regular (or symmetric) [8]. These are the square tessellation of the plane, Fig. 3, the triangular, Fig. 4, and the hexagonal or honeycomb tessellation, Fig. 5. The central node in Fig. 2 is included in three distinct planes of the square tessellation and four distinct planes of the triangular tessellation, which also includes the honeycomb tessellation.

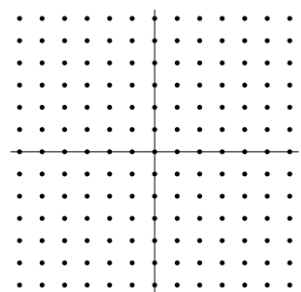

**Figure 3**: Square tessellation of the plane.

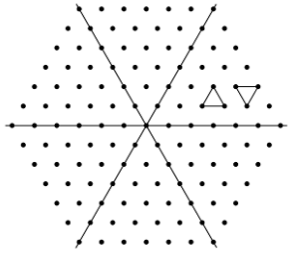

**Figure 4**: Triangular tessellation.

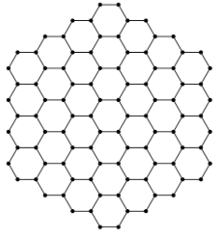

**Figure 5**: Honeycomb tessellation.

Fig. 6 illustrates a nucleated cube on the Bell Fuller lattice. Here the term 'nucleated' means that there is a node at the centre and there are nodes at each of the vertices of the cube. This cube contains a nucleated regular octahedron, whose vertices are the centre points of the six faces of the cube. It also contains two intersecting, nucleated, regular tetrahedra whose vertices are the vertices of the cube. These are the most compact nucleated regular tetrahedra that can be constructed on the lattice. It can be seen from this diagram that nucleated cubes based on the Bell Fuller lattice are both face-centred and body-centred. All red links shown in the lattice have the same length. Taking this as the unit length, the edge of the cube shown measures $2\sqrt{2}$, the face diagonal measures 4 and the body diagonal measures $2\sqrt{6}$.

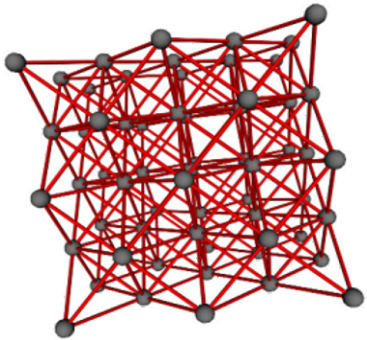

**Figure 6**: Nucleated cube on the Bell Fuller lattice.

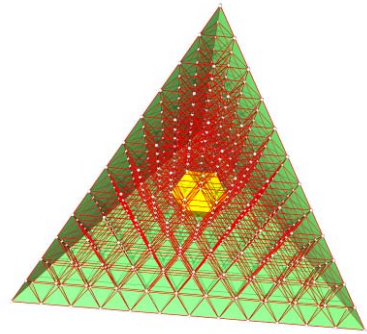

**Figure 7**: A nucleated tetrahedron on the Bell Fuller lattice.

Fig. 7 illustrates a nucleated tetrahedron of edge length 12 constructed on the Bell Fuller lattice. In this diagram there is a nucleated cuboctahedron shown at the centre of volume of the tetrahedron. All these lattice diagrams were prepared with the aid of a computer program called Springdance [9] and were rendered using POV-Ray [10].

If spacetime were a lattice of discrete points then this beautiful lattice, the Bell Fuller lattice, would be a strong candidate to represent its shape.

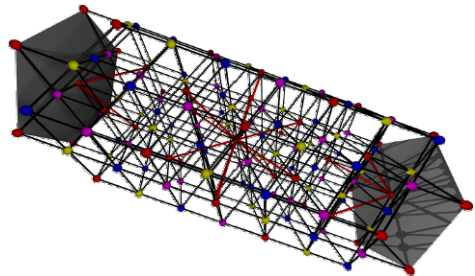

**Figure 8**: Diagram of points from four interlaced Bell Fuller lattices. (The end faces are regular hexagons and are parallel.)

## 4. FOUR BELL FULLER LATTICES

The points or nodes of precisely four Bell Fuller lattices of different colours can be interlaced in a symmetrical manner, as illustrated in Fig. 8. Any one node is surrounded by eight nearest neighbours: four of each of two of the other colours, e.g. red has four nearest neighbours that are yellow and four that are violet. The underlying lattice, with black links in Fig. 8, is *unique* in that its cells are all identical rhombohedra and it is the composition of four Bell Fuller lattices. It is proposed here that this four-colour rhombohedral lattice is the prime candidate to represent discrete spacetime, the vacuum.

Fig. 9 shows the same model as in Fig. 8 with the 'points' inflated to spheres that just touch. This is not to imply that the vacuum is made up of spheres: points have no radius. Figs. 10, 11 and 12 illustrate a set of points that form a rhombic dodecahedron on the four-colour rhombohedral lattice.

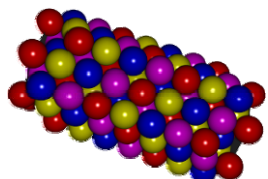

**Figure 9**: Inflated 'points' of the four-colour rhombohedral lattice.

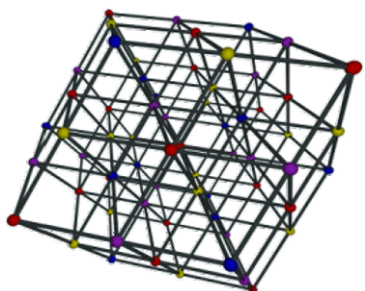

**Figure 10**: Rhombic dodecahedron on the four-colour rhombohedral lattice.

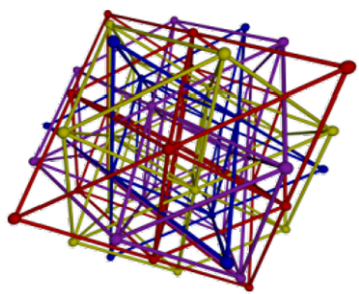

**Figure 11**: Links of the four Bell Fuller lattices that constitute the four-colour rhombohedral lattice.

Every point position on the four-colour rhombohedral lattice can be described by an integer 4-tuple. This is a 4-dimensional space. It possesses many, if not all, of the types of symmetry found in particle physics, quantum electrodynamics and quantum chromodynamics. The full range of the set of integers can be used arbitrarily within the 4-tuple to describe the position of any point with respect to a specified 'origin' and therefore the representation is not unique. However, any such 4-tuple can be converted to zero-based form without negative integers by the addition of a unique integer to all elements that makes at least one of the elements zero and all elements non-negative. A 'standard form' 4-tuple can also likewise be defined in which at least one of the elements is unity and none are zero or negative. This appears (to the present author) to be a more natural representation, with the origin described as (1, 1, 1, 1). Maintaining one ordinate at unity while three vary is equivalent to 'travelling' with a moving origin, as we do with an origin on the surface of the Earth.

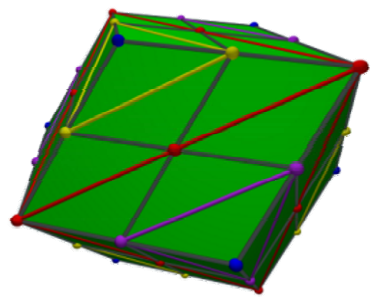

**Figure 12**: Convex hull of the points shown in Figs. 10 and 11 (a rhombic dodecahedron).

Any plane of the lattice of Fig. 10 that is made up of rhombs of unit edge length has an invariant metric ground form (see [3], p. 97) that allows the length of a vector to be defined. This concept can be extended to the four-dimensional discrete space of the lattice. The application and implementation of this approach (including conversions between 4-tuple and Cartesian coordinates) has been described in [11], [12], and [13], in the latter two cases using the term 'quadrays' for the four basis vectors. The 'length' formula for a vector, which can be derived using the cosine rule, is:

$$|\mathbf{r}| = \sqrt{\left(x_1^2 + x_2^2 + x_3^2 + x_4^2\right) - \frac{2}{3}\left(\begin{array}{l}x_1x_2 + x_2x_3 + x_3x_4 + \\ x_4x_1 + x_3x_1 + x_2x_4\end{array}\right)}$$

For example, the vector that goes from (1, 2, 1, 1) to (2, 1, 1, 1) is (2, 1, 1, 1) – (1, 2, 1, 1) = (1, –1, 0, 0), or (3, 1, 2, 2) in standard form. From the above formula it has a length of 2√(2/3), measured in quadray units.

The four basis quadrays of the four-colour rhombohedral lattice point to four of the corners of the smallest cube that can be constructed on the lattice, which has an edge length of 2/√3 quadray units. However, there are cubes of this size centred on each interior node and there are also cubes that have a vertex at the node.

Every point on the four-colour rhombohedral (or quadray) lattice can be mapped to a point on a conventional Cartesian 3-D lattice with integer ordinates (positive and negative) where the base distance is 1/√3 quadray units. The reverse is not true. Going in this direction, non-integer rational ordinates with a denominator of 2 may be produced, but only integer ordinates are permitted on the discrete quadray lattice. The nearest integer ordinate could be substituted, but thus there would be 'uncertainty' in the position on the quadray lattice. This is remarkably suggestive of the Heisenberg uncertainty principle. For instance, (1, –2, 3) maps onto (3, 1, 7/2, 5/2), which could be 'rounded up' to (3, 1, 4, 3). The latter 4-tuple would map back to (1, –3, 3). However, (3, 1, 7/2, 5/2) could just as well be 'rounded' to (3, 1, 3, 2), (3, 1, 4, 2), or (3, 1, 3, 3), all with equivalent correctness or probability.

There is a resolution to the 'apparent' uncertainty of position when referring to the points of the quadray lattice using integer Cartesian coordinates: integer Cartesian coordinate lists containing a mixture of odd and even integers are disallowed for purely geometric reasons: the quadray lattice maps to Cartesian integer coordinates where the integers in each coordinate list (3-tuple) are either even or odd, but not a mixture of both.

## 5. TIME AND DISTANCE

In 1879 in a classic experiment of science Albert Michelson [14] measured the speed of light (in air) to very high precision. Also, to date, the speed of light in a vacuum has been found to be invariant. An alternative interpretation of Michelson's experiment could be that he measured the 'distance equivalent of time', just as Joule measured the 'mechanical equivalent of heat'. As corroborating evidence in support of this interpretation, distance is now formally defined, since 1983, in terms of time [15]. The speed of light may be no more than an *integer* conversion factor between different possible units for measuring spacetime, namely the metre and the second.

$$299\ 792\ 458\ [m] = 1[s]$$

In coherent units the velocity of light has a value of unity. These statements about a fundamental physical constant are not as radical as they may seem and the general concept is not new. Particle physicists commonly work with such units [16]. The light-year used by cosmologists is the distance-equivalent of one year of time. Time intervals and distance intervals can both be expressed in light-years, or even light-nanoseconds.

## 6. CHANGE, MOTION AND ACTION

The existence of a pattern other than the spacetime lattice itself requires that each point of a given colour can have at least two states, e.g. light and dark. Anything that one might consider a 'particle' exists not just at one point in spacetime, but as a repeating pattern (of perhaps very many points) along a lattice path (or string) in spacetime. If there is acceleration then the patterns repeat on a 'curved path' or at decreasing intervals.

Change is a morphing transition between patterns in spacetime. The motion of a pattern can be quantified provided the pattern remains recognisable over a path in spacetime. If a string of arbitrary thickness is drawn in spacetime and if an identifiable pattern is found to repeat periodically within it over its length, the speed of that pattern is the number of repetitions of the pattern divided by the length of the string. The length of the string is the integer count of the number of lattice points its 'centre' passes over. The framework of spacetime (the vacuum) is itself a pattern that repeats at every lattice point. Therefore the framework itself has a speed of unity, or $c$ m/s, where $c$ is the 'distance equivalent of time'. No other pattern can repeat more 'quickly'.

Relative movement or action can be explained in terms of patterns too. In Fig. 13, Pattern 1 has a period of 12, whereas Pattern 2 has a period of 14. If Pattern 1, with a sequence from left to right, is the 'observer' then Pattern 2 moves to the right with a relative 'speed' of 2/12. If Pattern 2, also with a sequence from left to right, is 'observer', then Pattern 1 moves to the left with a relative speed of 2/14. Assuming the patterns are collinear, the relative speeds mentioned are both constant.

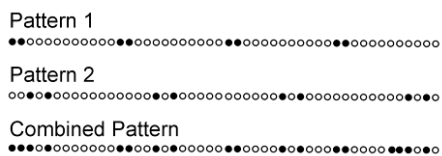

**Figure 13**: Relative movement of patterns.

## 7. THERMODYNAMICS

Engineering thermodynamicists find it convenient to work with the concept of 'specific energy', or energy per unit mass. There is a 'double-think' involved when one considers that mass is energy. In fact if spacetime is fully symmetric then, from Newton's second law, force is mass over distance. It would seem that mass (or energy) itself is measured in distance units and quantifies the size of a particle pattern about a point.

In general relativity the time-order in which two events occur is dependent on the observer. This must also necessarily be the case in thermodynamics. A clock is a device (itself a repeating pattern) that registers being at a spacetime point of a particular type (e.g. red, yellow, blue, violet). An observer travelling with the clock can keep a count of the distance travelled. Prima facie it is not inconceivable that all Einstein's relativistic effects, which are massively supported by experimental evidence, are consistent with pointwise discrete spacetime, based on the four-colour rhombohedral lattice. If so, some major simplifications of fundamental physical constants should result.

Lieb and Yngvason [17] have presented a well founded thermodynamics theory based on the ordering of sets. It includes the concept of entropy and that of temperature derived from it. In the areas of cosmology and particle physics the Hawking temperature [18] (the temperature at which a black hole radiates energy) and the Unruh temperature (the temperature to which an accelerated observer is excited while travelling through the vacuum) are both proportional to acceleration (in the former case this is the acceleration due to gravity at the surface of the black hole). All these thermodynamic concepts are geometric in nature. They depend on the shape of spacetime itself and on the shape of the path travelled by an observer. They depend on symmetry. Mathematicians and physicists have clearly established the relationship between conservation laws and the symmetry of groups [19]. However, it is important to point out that the discrete lattices described here involve discrete groups rather than Lie groups, which are continuous.

## 8. CONCLUSION

If the vacuum of flat spacetime were a pointwise 4-colour rhombohedral lattice then spacetime, shape and symmetry would have everything to do with thermodynamics.